\documentclass[10pt,twoside,english]{article}
\usepackage[T1]{fontenc}
\usepackage[latin9]{inputenc}
\usepackage[a4paper]{geometry}
\usepackage{amsmath}
\usepackage{amssymb}
\usepackage{graphicx}
\usepackage{setspace}
\usepackage{esint}
\usepackage{babel}
\begin{document}

\title{Notes on steady state current through discrete-level quantum systems}

\author{Longwen Zhou\thanks{zhoulw13@u.nus.edu}}
\maketitle
\begin{abstract}
In these notes, we take a naive approach to calculate electrical current
through a noninteracting quantum system with discrete energy levels.
We do not assume any prior knowledge on second quantization, scattering
matrix and nonequilibrium Green's functions (NEGF). Instead, we will
try to build our solutions to the problem step by step from single-particle
Schrodinger equation and equilibrium Green's functions. In the \textbf{first
section}, we give the definitions of retarded Green's function, spectral
function and density of states for time-independent quantum systems.
In the \textbf{second section}, we introduce the left-center-right
(LCR) system, which may be viewed as a minimum model for the study
of quantum transport. In the \textbf{third section}, we work out the
Green's function and spectral function of the central part in the
LCR system. Important concepts like self-energy and level-width function
will also be introduced. In the \textbf{fourth section}, we will derive
the Landauer formula for steady state current using a scattering approach.
The transmission coefficient at a given energy will be derived. In
the \textbf{fifth section}, we will apply Landauer formula to the
case in which the left and right leads are both semi-infinite tight-binding
chains, coupling to the central system only at its boundaries. In
this case, further simplifications can be made under wide-band limit.
In the \textbf{sixth section}, we take the zero temperature, zero
bias limit to obtain the linear conductance from Landauer formula.
In the \textbf{seventh section}, we apply the linear conductance formula
under wide-band limit to two typical models of the central system:
a single-level quantum dot and a double-level quantum dot. In both
cases, we will calculate the spectral function and electrical conductance
analytically. The results illustrate how quantum resonance and coherence
can affect electrical transport in noninteracting systems. In the
\textbf{eighth section}, we use the Landauer formula to compute linear
conductance of a tight-binding chain. A simple example of metal-insulator
transition is discussed using the Aubry-André-Harper model. In the
\textbf{ninth section}, Landauer formula is applied to study edge
state transport along the boundary of a two-dimensional lattice. Using
the Hofstadter model as an example, we show the quantization of linear
conductance in the spectral gap, which indicates the topological nontrivial
properties of the system. In the \textbf{last section}, we will give
a summary and discuss possible future extensions. 

The first seven sections of these notes follow closely Chapter 3 of
Ref.\cite{RyndykBook}. Other materials guiding the preparation of
these notes are Chapter 13 of Ref.\cite{JishiBook}, Chapter 3 of
Ref.\cite{VentraBook}, the monograph \cite{CuevasBook} and the review
paper \cite{WangNEGFRev2014}.

\newpage{}
\end{abstract}
\tableofcontents{}

\newpage{}

\section{Retarded Green's function, spectral function and density of states}

The \textbf{retarded Green's function (or propagator)} of a time-independent
quantum system described by Hamiltonian $H$ is defined as
\begin{equation}
G(t,t')=-\frac{i}{\hbar}\theta(t-t')e^{-\frac{i}{\hbar}H(t-t')},\label{eq:GRt}
\end{equation}
where $\hbar$ is the Planck constant, $t$ and $t'$ are time variables.
The step function $\theta(t-t')$ is defined as
\begin{equation}
\theta(t-t')=\begin{cases}
1 & t\geq t'\\
0 & t<t'
\end{cases}.\label{eq:step}
\end{equation}
The retarded Green's function is the solution of the \textbf{equation
of motion}:
\begin{equation}
\left(i\hbar\frac{\partial}{\partial t}-H\right)G(t,t')=\delta(t-t')\label{eq:eom}
\end{equation}
with boundary condition
\begin{equation}
i\hbar\lim_{\eta\rightarrow0^{+}}G(t'+\eta,t')=1,\label{eq:eom-bc}
\end{equation}

The \textbf{Fourier transform} of $G(t,t')$ from time to energy domain
is given by
\begin{equation}
G(E)=\int_{-\infty}^{\infty}dEe^{\frac{i}{\hbar}(E+i0^{+})(t-t')}G(t,t'),\label{eq:GR-ft}
\end{equation}
where $0^{+}$ represents an infinitesimal positive number. The Fourier
expansion of $\delta(t-t')$ reads
\begin{equation}
\delta(t-t')=\frac{1}{2\pi\hbar}\int_{-\infty}^{\infty}e^{-\frac{i}{\hbar}E(t-t')}dE.\label{eq:Delta-ft}
\end{equation}
Combining Eqs. (\ref{eq:eom}), (\ref{eq:GR-ft}) and (\ref{eq:Delta-ft}),
we obtain the \textbf{retarded Green's function in energy domain}
as
\begin{equation}
G(E)=\left(E+i0^{+}-H\right)^{-1}.\label{eq:GRe}
\end{equation}
Note that $G(E)$ here has the dimension of $E^{-1}$. Using $G(E)$,
we can introduce the \textbf{spectral operator} as
\begin{equation}
{\bf A}(E)=i\left[G(E)-G^{\dagger}(E)\right]=2\pi\delta(E-H),\label{eq:spectralm}
\end{equation}
where we have used the \textbf{Plemelj formula} $\frac{1}{x\pm i0^{+}}={\cal P}\frac{1}{x}\mp i\pi\delta(x)$,
with ${\cal P}$ standing for the Cauchy principle value. The \textbf{spectral
function} $A$ and \textbf{density of states} $\varrho$ at a given
energy $E$ are obtained from the spectral operator as:
\begin{alignat}{1}
A(E) & ={\rm Tr}[{\bf A}(E)],\label{eq:spectralf}\\
\varrho(E) & =\frac{1}{2\pi}A(E).\label{eq:dos}
\end{alignat}
Note that the trace is taken in the Hilbert space of Hamiltonian $H$.
From now on we will work in energy domain only.

\section{LCR system}

An \textbf{LCR system} may be regarded as the minimum model for the
study of quantum transport. It is usually adopted in the description
of electrical current through a noninteracting quantum dot or molecule
junction. In matrix form, the Hamiltonian of an LCR system can be
expressed as
\begin{equation}
H=\begin{pmatrix}H_{LL} & H_{LC} & 0\\
H_{CL} & H_{CC} & H_{CR}\\
0 & H_{RC} & H_{RR}
\end{pmatrix},\label{eq:Hlcr}
\end{equation}
where $H_{LL}$ and $H_{RR}$ are Hamiltonians of left $(L)$ and
right $(R)$ leads, respectively. $H_{CC}$ is the Hamiltonian of
the central $(C)$ region, whose transport property is of our interest.
$H_{LC}=H_{CL}^{\dagger}$ describes the coupling between the central
region and the left lead, and $H_{RC}=H_{CR}^{\dagger}$ describes
the coupling between the central region and the right lead. There
is no direct coupling between left and right leads. The eigenvalue
equation of Hamiltonian $H$ is given by
\begin{equation}
\begin{pmatrix}H_{LL} & H_{LC} & 0\\
H_{CL} & H_{CC} & H_{CR}\\
0 & H_{RC} & H_{RR}
\end{pmatrix}\begin{pmatrix}\Psi_{L}\\
\Psi_{C}\\
\Psi_{R}
\end{pmatrix}=E\begin{pmatrix}\Psi_{L}\\
\Psi_{C}\\
\Psi_{R}
\end{pmatrix}\label{eq:Seq-lcr}
\end{equation}
where the wave function components $\Psi_{L},\Psi_{C}$ and $\Psi_{R}$
are written in a basis well-localized in each of the three regions
$L,C$ and $R$.

\section{Green's function and spectral function of an LCR system}

The retarded Green's function of an LCR system can also be written
in matrix form as

\begin{equation}
G=\begin{pmatrix}G_{LL} & G_{LC} & G_{LR}\\
G_{CL} & G_{CC} & G_{CR}\\
G_{RL} & G_{RC} & G_{RR}
\end{pmatrix}.\label{eq:GRlcr}
\end{equation}
It satisfies Eq. (\ref{eq:GRe}) with $H$ given by Eq. (\ref{eq:Hlcr}):
\begin{equation}
\begin{pmatrix}E-H_{LL}\pm i0^{+} & -H_{LC} & 0\\
-H_{CL} & E-H_{CC}\pm i0^{+} & -H_{CR}\\
0 & -H_{RC} & E-H_{RR}\pm i0^{+}
\end{pmatrix}\begin{pmatrix}G_{LL} & G_{LC} & G_{LR}\\
G_{CL} & G_{CC} & G_{CR}\\
G_{RL} & G_{RC} & G_{RR}
\end{pmatrix}=1.
\end{equation}
The Green's function of the central region $G_{CC}$ is then determined
by the following set of equations:
\begin{alignat}{1}
G_{LC}= & G_{LL}^{0}H_{LC}G_{CC}\\
(E-H_{CC}\pm i0^{+})G_{CC}= & H_{CL}G_{LC}+H_{CR}G_{RC}\\
G_{RC}= & G_{RR}^{0}H_{RC}G_{CC},
\end{alignat}
where we have introduced Green's functions of isolated left and right
leads as
\begin{equation}
G_{\alpha\alpha}^{0}=(E-H_{\alpha\alpha}\pm i0^{+})^{-1}\qquad\alpha=L,R.\label{eq:GRleads}
\end{equation}
Solving these equations gives us
\begin{equation}
G_{CC}=\frac{1}{E-H_{CC}-\Sigma_{L}-\Sigma_{R}+i0^{+}},\label{eq:GRcenter}
\end{equation}
where the \textbf{retarded self-energy} functions $\Sigma_{L,R}$
of the two leads are:
\begin{equation}
\Sigma_{\alpha}\equiv H_{C\alpha}G_{\alpha\alpha}^{0}H_{\alpha C}\qquad\alpha=L,R.\label{eq:self-eng}
\end{equation}
The self-energy functions incorporate all effects of the lead on the
central system. One may combine $H_{CC}$ and the self-energies to
obtain an \textbf{effective Hamiltonian} for the ``\textbf{dressed}''
central region:
\begin{equation}
H_{CC}^{{\rm eff}}\equiv H_{CC}+\Sigma_{L}+\Sigma_{R}.\label{eq:HCeff}
\end{equation}
However, this Hamiltonian is \textbf{non-Hermitian} and its spectrum
is in general not real. Therefore it does not describe a closed quantum
system. The open system nature of $H_{CC}^{{\rm eff}}$ reflects the
fact that all degrees of freedom of the lead have been integrated
out in order to obtain the self-energies. For $H_{CC}^{{\rm eff}}$,
the real part of its spectrum reflects the energy \textbf{level shift}
caused by coupling to the leads, and the imaginary part of its spectrum
determines the \textbf{lifetime} of the ``dressed'' energy level.
As can be inspected from Eq. (\ref{eq:self-eng}), the lifetime of
a ``dressed'' energy level is in general proportional to the inverse
square of the coupling strength between the lead and the central region.
To characterize the \textbf{level-broadening} caused by system-lead
coupling, we introduce the \textbf{level width function} as ($\alpha=L,R$):
\begin{equation}
\Gamma_{\alpha}\equiv i\left(\Sigma_{\alpha}-\Sigma_{\alpha}^{\dagger}\right)=H_{C\alpha}i\left(G_{\alpha\alpha}^{0}-G_{\alpha\alpha}^{0\dagger}\right)H_{\alpha C}=2\pi H_{C\alpha}\delta(E-H_{\alpha\alpha})H_{\alpha C}.\label{eq:level-width}
\end{equation}
We see that the $\Gamma_{\alpha}$ is proportional to the density
of states of lead $\alpha$ and the square of the coupling strength
between the lead $\alpha$ and the central system.

Finally, the spectral function and density of states of the central
region are defined following Eq. (\ref{eq:spectralm}) to (\ref{eq:dos})
as:
\begin{alignat}{1}
{\bf A}_{C}(E) & =i\left[G_{CC}(E)-G_{CC}^{\dagger}(E)\right],\label{eq:spectralm-cc}\\
A_{C}(E) & ={\rm Tr}[{\bf A}_{C}(E)],\label{eq:spectralf-cc}\\
\varrho_{C}(E) & =\frac{1}{2\pi}A_{C}(E).\label{eq:dos-cc}
\end{alignat}
With all these preparations, we will derive Landauer formula for steady
state transport using a scattering approach in the next section.

\section{Landauer formula and transmission coefficient}

We will take a \textbf{scattering point of view} to describe the charge
transport in our system. In all the derivations below, we assume there
is no many-body interactions. We start with the stationary Schrodinger
equation of the LCR system:
\begin{equation}
\begin{pmatrix}H_{LL} & H_{LC} & 0\\
H_{CL} & H_{CC} & H_{CR}\\
0 & H_{RC} & H_{RR}
\end{pmatrix}\begin{pmatrix}\Psi_{L}\\
\Psi_{C}\\
\Psi_{R}
\end{pmatrix}=E\begin{pmatrix}\Psi_{L}\\
\Psi_{C}\\
\Psi_{R}
\end{pmatrix}.\label{eq:Seq-lcr-1}
\end{equation}
Consider an incoming wave $\Psi_{L}^{0}$ from the left the lead to
the central region, which is an eigenstate of $H_{LL}$. This wave
may be partially transmitted into the central region, and partially
reflected back to the left lead, yielding a reflection wave $\Psi_{L}^{1}$.
Therefore for such an incoming wave, we can write the wave function
in the left lead as
\begin{equation}
\Psi_{L}=\Psi_{L}^{0}+\Psi_{L}^{1}.\label{eq:wave-left}
\end{equation}
Plugging Eq. (\ref{eq:wave-left}) into Eq. (\ref{eq:Seq-lcr-1})
gives us
\begin{alignat}{1}
(E-H_{LL})(\Psi_{L}^{0}+\Psi_{L}^{1}) & =H_{LC}\Psi_{C},\\
(E-H_{CC})\Psi_{C} & =H_{CL}(\Psi_{L}^{0}+\Psi_{L}^{1})+H_{CR}\Psi_{R},\\
(E-H_{RR})\Psi_{R} & =H_{RC}\Psi_{C}.
\end{alignat}
Using the eigenvalue equation $(E-H_{LL})\Psi_{L}^{0}=0$, definitions
of lead Green's functions Eq. (\ref{eq:GRleads}) and self-energies
Eq. (\ref{eq:self-eng}), we obtain the following expressions for
different wave function components in the left lead, central region
and right lead:
\begin{alignat}{1}
\Psi_{L} & =(1+G_{LL}^{0}H_{LC}G_{CC}H_{CL})\Psi_{L}^{0},\label{eq:PsiL}\\
\Psi_{C} & =G_{CC}H_{CL}\Psi_{L}^{0},\label{eq:PsiC}\\
\Psi_{R} & =G_{RR}^{0}H_{RC}G_{CC}H_{CL}\Psi_{L}^{0}.\label{eq:PsiR}
\end{alignat}

When a steady state is established, the probability of the state in
the central region $|\Psi_{C}|^{2}$ should not change with time.
This is the case if the current approaching the central region equals
the current leaving it. The definition of steady state current can
then be extracted from probability conservation law as follows \cite{PaulssonNotes}:
\begin{alignat}{1}
0=\frac{d\Psi_{C}^{\dagger}\Psi_{C}}{dt}= & \Psi_{C}^{\dagger}\frac{d\Psi_{C}}{dt}+\frac{d\Psi_{C}^{\dagger}}{dt}\Psi_{C}\nonumber \\
= & \Psi_{C}^{\dagger}\frac{1}{i\hbar}(H_{CL}\Psi_{L}+H_{CC}\Psi_{C}+H_{CR}\Psi_{R})\nonumber \\
- & \frac{1}{i\hbar}(\Psi_{L}^{\dagger}H_{LC}+\Psi_{C}^{\dagger}H_{CC}+\Psi_{R}^{\dagger}H_{RC})\Psi_{C}\nonumber \\
= & \frac{1}{i\hbar}(\Psi_{C}^{\dagger}H_{CL}\Psi_{L}-\Psi_{L}^{\dagger}H_{LC}\Psi_{C})\\
+ & \frac{1}{i\hbar}(\Psi_{C}^{\dagger}H_{CR}\Psi_{R}-\Psi_{R}^{\dagger}H_{RC}\Psi_{C}).
\end{alignat}
We can now interpret the (local) \textbf{probability current} from
lead $\alpha$ to the central region as
\begin{equation}
J_{\alpha}\equiv\frac{1}{i\hbar}(\Psi_{C}^{\dagger}H_{C\alpha}\Psi_{\alpha}-\Psi_{\alpha}^{\dagger}H_{\alpha C}\Psi_{C})\qquad\alpha=L,R.\label{eq:ProbCurrJ}
\end{equation}
The conservation law is then given by
\begin{equation}
\sum_{\alpha=L,R}J_{\alpha}=0\Leftrightarrow J_{L}=-J_{R}.
\end{equation}

Using Eq. (\ref{eq:ProbCurrJ}) and Eqs. (\ref{eq:PsiL}) to (\ref{eq:PsiR}),
we can compute the contribution of the incoming wave $\Psi_{L}^{0}$
to the current from the left lead to the central region as:
\begin{alignat}{1}
J_{L}= & \frac{i}{\hbar}(\Psi_{C}^{\dagger}H_{CR}\Psi_{R}-\Psi_{R}^{\dagger}H_{RC}\Psi_{C})\nonumber \\
= & \frac{i}{\hbar}\Psi_{L}^{0\dagger}H_{LC}G_{CC}^{\dagger}(H_{CR}G_{RR}^{0}H_{RC}-H_{CR}G_{RR}^{0\dagger}H_{RC})G_{CC}H_{CL}\Psi_{L}^{0}\nonumber \\
= & \frac{i}{\hbar}\Psi_{L}^{0\dagger}H_{LC}G_{CC}^{\dagger}(\Sigma_{R}-\Sigma_{R}^{\dagger})G_{CC}H_{CL}\Psi_{L}^{0}\nonumber \\
= & \frac{1}{\hbar}\Psi_{L}^{0\dagger}H_{LC}G_{CC}^{\dagger}\Gamma_{R}G_{CC}H_{CL}\Psi_{L}^{0},
\end{alignat}
where we have used our definitions of self-energy and level width
function given by Eq. (\ref{eq:self-eng}) and Eq. (\ref{eq:level-width})
in the last section. To proceed, we assume that all possible incoming
states are independent and originated from the same Fermi function
of the left lead $f_{L}(\epsilon_{kL})$, where $\epsilon_{kL}$ is
the energy of incoming state $\Psi_{kL}^{0}$ with quantum number
$k$. Also we assume that there is no scattering among different incoming
channels. Under these conditions, the total probability current from
the left lead to the central region is given by
\begin{alignat}{1}
I_{L}= & \sum_{k}\frac{1}{\hbar}\Psi_{kL}^{0\dagger}H_{LC}G_{CC}^{\dagger}\Gamma_{R}G_{CC}H_{CL}\Psi_{kL}^{0}f_{L}(\epsilon_{kL})\label{eq:IL}\\
= & \frac{1}{\hbar}\sum_{k}\sum_{q}\Psi_{kL}^{0\dagger}H_{LC}\Psi_{qC}^{0}\Psi_{qC}^{0\dagger}G_{CC}^{\dagger}\Gamma_{R}G_{CC}H_{CL}\Psi_{kL}^{0}f_{L}(\epsilon_{kL})\nonumber \\
= & \frac{1}{h}\sum_{q}\Psi_{qC}^{0\dagger}G_{CC}^{\dagger}\Gamma_{R}G_{CC}H_{CL}\left[2\pi\sum_{k}\Psi_{kL}^{0}\Psi_{kL}^{0\dagger}f_{L}(\epsilon_{kL})\right]H_{LC}\Psi_{qC}^{0}\nonumber \\
= & \frac{1}{h}\int_{-\infty}^{\infty}dE\sum_{q}\Psi_{qC}^{0\dagger}G_{CC}^{\dagger}\Gamma_{R}G_{CC}H_{CL}\left[2\pi\sum_{k}\delta(E-\epsilon_{kL})\Psi_{kL}^{0}\Psi_{kL}^{0\dagger}\right]H_{LC}\Psi_{qC}^{0}f_{L}(E),\nonumber 
\end{alignat}
where we have inserted the completeness relation $\sum_{q}\Psi_{qC}^{0}\Psi_{qC}^{0\dagger}=1_{C}$,
with $q$ being the quantum number of the central region. Recalling
Eq. (\ref{eq:level-width}) for the level-width function, we notice
that
\begin{equation}
H_{CL}\left[2\pi\sum_{k}\delta(E-\epsilon_{kL})\Psi_{kL}^{0}\Psi_{kL}^{0\dagger}\right]H_{LC}=\Gamma_{L}(E).
\end{equation}
Combining this observation with Eq. (\ref{eq:IL}) yields
\begin{alignat}{1}
I_{L}= & \frac{1}{h}\int_{-\infty}^{\infty}dE\sum_{q}\Psi_{qC}^{0\dagger}G_{CC}^{\dagger}\Gamma_{R}G_{CC}^{r}\Gamma_{L}\Psi_{qC}^{0}f_{L}(E)\nonumber \\
= & \frac{1}{h}\int_{-\infty}^{\infty}dE{\rm Tr}\left(G_{CC}^{\dagger}\Gamma_{R}G_{CC}\Gamma_{L}\right)f_{L}(E),
\end{alignat}
where the trace is over localized basis of the central region. Finally,
taking into account the contributions from both left and right leads,
the total steady state probability current $I=I_{L}+I_{R}$ flowing
into the central region is given by:
\begin{equation}
I=\frac{1}{h}\int_{-\infty}^{\infty}dET(E)\left[f_{L}(E)-f_{R}(E)\right],\label{eq:Landauer}
\end{equation}
where the transmission coefficient $T(E)$ is defined as
\begin{equation}
T(E)\equiv{\rm Tr}\left[G_{CC}^{\dagger}(E)\Gamma_{R}(E)G_{CC}(E)\Gamma_{L}(E)\right].\label{eq:Transcoeff}
\end{equation}
Eq. (\ref{eq:Landauer}) is the \textbf{Landauer formula} for steady
state transport in noninteracting quantum systems. The current is
determined by the transmission coefficient times the difference of
electron distributions in left and right leads at a given energy $E$,
and integrating over all possible energies of incoming state. In this
picture, we can roughly say that current is transmission. Other popular
approaches to the derivation of Landauer formula including the scattering
matrix formalism and NEGF. Interested readers can consult Refs.\cite{JishiBook,VentraBook}
for further details.

In the next section, we will try to obtain a more explicit expression
for $T(E)$ by specifying the Hamiltonians of lead and coupling between
leads and central region.

\section{Semi-infinite tight-binding leads and wide-band limit}

Our derivations up to now are formal. In the following, we will specify
the Hamiltonians of lead and system-lead coupling in order to arrive
at a more explicit expression for the transmission coefficient. The
Hamiltonians for the lead and their coupling to the central region
may be written quite generally as (for $\alpha=L,R$):
\begin{alignat}{1}
H_{\alpha\alpha} & =\sum_{m,n}t_{\alpha,mn}|m_{\alpha}\rangle\langle n_{\alpha}|,\label{eq:H-lead}\\
H_{C\alpha} & =\sum_{\mu,n}t_{C\alpha,\mu n}|\mu\rangle\langle n_{\alpha}|=H_{\alpha C}^{\dagger}.\label{eq:H-coupling}
\end{alignat}
where $\{|m_{\alpha}\rangle\}$ is a localized basis of lead $\alpha$.
$\{|\mu\rangle\}$ is a localized basis of the central region, which
will be assumed to have a finite dimension $N$. 

Since in most cases, leads are just sources of electrons, we may simply
choose them to be semi-infinite tight-binding chains. Moreover, we
allow each basis $|\mu\rangle$ of the central region to be coupled
separately to one tight-binding chain, and require that all the tight-binding
chains coupled to the central region at the same side $\alpha$ are
equivalent. The Hamiltonians for lead may then be written as
\begin{alignat}{1}
H_{LL} & =\sum_{\mu=0}^{N-1}\left[t_{L}\sum_{m=-\infty}^{-2}(|m,\mu\rangle\langle m+1,\mu|+{\rm h.c.})+w_{L}\sum_{m=-\infty}^{-1}|m,\mu\rangle\langle m,\mu|\right],\label{eq:H-left-chain}\\
H_{RR} & =\sum_{\mu=0}^{N-1}\left[t_{R}\sum_{m=1}^{\infty}(|m,\mu\rangle\langle m+1,\mu|+{\rm h.c.})+w_{R}\sum_{m=1}^{\infty}|m,\mu\rangle\langle m,\mu|\right],\label{eq:H-right-chain}
\end{alignat}
where $t_{L}\geq0$ ($t_{R}\geq0$) is the nearest neighbor hopping
amplitude, and $w_{L}$ ($w_{R}$) is the onsite potential of the
left (right) tight-binding chain. The lattice constant has been set
to $1$, and each of the left (right) chain extends from site $-\infty$
($+1$) to site $-1$ ($+\infty$). The tensor product basis $\{|m,\mu\rangle\}$
is complete, and satisfies the orthonormal condition
\begin{equation}
\langle m,\mu|n,\nu\rangle=\delta_{mn}\delta_{\mu\nu}.\label{eq:orthonormal}
\end{equation}
Next, we note that for each $\mu$, the semi-infinite tight-binding
chain is a \textbf{tridiagonal Toeplitz matrices} in our basis. Therefore
we obtain the dispersion relation $\epsilon_{k_{\alpha}\mu}$ and
eigenfunctions $\{|k_{\alpha},\mu\rangle\}$ of the $\mu$'s tight-binding
chain as \cite{TTM}:
\begin{alignat}{1}
\epsilon_{k_{\alpha}\mu} & =2t_{\alpha}\cos(k_{\alpha})+w_{\alpha},\label{eq:lead-dispersion}\\
\langle m,\mu|k_{\alpha},\nu\rangle & =\sin(k_{\alpha}m)\delta_{\mu\nu},\label{eq:lead-eigenstate}
\end{alignat}
where $\alpha=L,R$ and $\mu=0,...,N-1$.

To evaluate the transmission coefficient $T(E)$, we need to calculate
self-energies and level width functions. Both of them require specific
knowledge of the couplings between leads and central system. Spatially,
the central system is located between sites $-1$ and $+1$. So we
model the couplings between lead and the central system as
\begin{alignat}{1}
H_{CL} & =\sum_{\mu=0}^{N-1}t_{\mu L}|\mu\rangle\langle-1,\mu|=H_{LC}^{\dagger},\label{eq:center-left-coupling}\\
H_{CR} & =\sum_{\mu=0}^{N-1}t_{\mu R}|\mu\rangle\langle+1,\mu|=H_{RC}^{\dagger},\label{eq:center-right-coupling}
\end{alignat}
where $t_{\mu\alpha}$ is the coupling strength between the $\mu$'s
basis of the central region and lead $\alpha$. Note that this coupling
is local in space. 

We are now ready to obtain a more explicit expression for transmission
coefficient $T(E)$ given by Eq. (\ref{eq:Transcoeff}). From Eq.
(\ref{eq:level-width}) and Eqs. (\ref{eq:orthonormal}) to (\ref{eq:center-right-coupling}),
the level-width function of lead $\alpha$ is given by
\begin{alignat}{1}
\Gamma_{\alpha}= & 2\pi H_{C\alpha}\delta(E-H_{\alpha\alpha})H_{\alpha C}\nonumber \\
= & 2\pi\sum_{k_{\alpha},\mu}\delta(E-\epsilon_{k_{\alpha}\mu})H_{C\alpha}|k_{\alpha},\mu\rangle\langle k_{\alpha},\mu|H_{\alpha C}\nonumber \\
= & 2\pi\sum_{k_{\alpha},\mu}|t_{\mu\alpha}|^{2}\sin^{2}(k_{\alpha})\delta(E-\epsilon_{k_{\alpha}\mu})|\mu\rangle\langle\mu|.
\end{alignat}
In the case that the lead $\alpha$ has a \textbf{continuous spectrum},
we can transform the sum over quantum number $k_{\alpha}$ to an integral:
\begin{alignat}{1}
\Gamma_{\alpha}= & \sum_{\mu}|t_{\mu\alpha}|^{2}|\mu\rangle\langle\mu|\int_{-\pi}^{\pi}\sin^{2}(k_{\alpha})\delta(E-\epsilon_{k_{\alpha}\mu})dk_{\alpha}\nonumber \\
= & \sum_{\mu}2|t_{\mu\alpha}|^{2}|\mu\rangle\langle\mu|\int_{0}^{\pi}\sin^{2}(k_{\alpha})\delta(E-\epsilon_{k_{\alpha}\mu})dk_{\alpha}\nonumber \\
= & \sum_{\mu}2|t_{\mu\alpha}|^{2}|\mu\rangle\langle\mu|\int_{0}^{\pi}\sin^{2}(k_{\alpha})\delta(E-\epsilon_{k_{\alpha}\mu})d\arccos\left(\frac{\epsilon_{k_{\alpha}\mu}-w_{\alpha}}{2t_{\alpha}}\right)\nonumber \\
= & \sum_{\mu}\frac{|t_{\mu\alpha}|^{2}}{t_{\alpha}}|\mu\rangle\langle\mu|\int_{w_{\alpha}-2t_{\alpha}}^{w_{\alpha}+2t_{\alpha}}\frac{\sin^{2}\left[\arccos\left(\frac{\epsilon_{k_{\alpha}\mu}-w_{\alpha}}{2t_{\alpha}}\right)\right]}{\sqrt{1-\left(\frac{\epsilon_{k_{\alpha}\mu}-w_{\alpha}}{2t_{\alpha}}\right)^{2}}}\delta(E-\epsilon_{k_{\alpha}\mu})d\epsilon_{k_{\alpha}\mu}\nonumber \\
= & \sum_{\mu}\frac{|t_{\mu\alpha}|^{2}}{t_{\alpha}}\frac{\sin^{2}\left[\arccos\left(\frac{E-w_{\alpha}}{2t_{\alpha}}\right)\right]}{\sqrt{1-\left(\frac{E-w_{\alpha}}{2t_{\alpha}}\right)^{2}}}|\mu\rangle\langle\mu|.
\end{alignat}
We note that due to our choices of the center-lead coupling, $\Gamma_{\alpha}$
is diagonal in the basis of the central region $\{|\mu\rangle\}$.
If the band width $2t_{\alpha}$ of the lead $\alpha$ is much wider
then $E-w_{\alpha}$, we can take the so-called \textbf{wide-band
limit}, in which we send $\frac{E-w_{\alpha}}{2t_{\alpha}}\rightarrow0$.
In this limit, we have $\sin^{2}\left[\arccos\left(\frac{E-w_{\alpha}}{2t_{\alpha}}\right)\right]\rightarrow1$
and the level-width function $\Gamma_{\alpha}$ becomes independent
of energy $E$. We will denote the level width function in wide-band
limit as
\begin{equation}
\Gamma_{\alpha}^{w}=\sum_{\mu=0}^{N-1}\frac{|t_{\mu\alpha}|^{2}}{t_{\alpha}}|\mu\rangle\langle\mu|\qquad\alpha=L,R.\label{eq:level-width-wide-band}
\end{equation}

In the same limit, the self-energies can also be worked out explicitly.
From Eq. (\ref{eq:self-eng}) and Eqs. (\ref{eq:orthonormal}) to
(\ref{eq:center-right-coupling}), we get
\begin{alignat}{1}
\Sigma_{\alpha}= & \sum_{k_{\alpha},\mu}\frac{|t_{\mu\alpha}|^{2}\sin^{2}(k_{\alpha})}{E-\epsilon_{k_{\alpha}\mu}+i0^{+}}|\mu\rangle\langle\mu|\nonumber \\
= & \frac{1}{2\pi}\sum_{\mu}|t_{\mu\alpha}|^{2}|\mu\rangle\langle\mu|\int_{-\pi}^{\pi}\frac{\sin^{2}(k_{\alpha})}{E-\epsilon_{k_{\alpha}\mu}+i0^{+}}dk_{\alpha}\nonumber \\
= & \frac{1}{\pi}\sum_{\mu}|t_{\mu\alpha}|^{2}|\mu\rangle\langle\mu|\int_{0}^{\pi}\frac{\sin^{2}(k_{\alpha})}{E-\epsilon_{k_{\alpha}\mu}+i0^{+}}dk_{\alpha}\nonumber \\
= & \frac{1}{\pi}\sum_{\mu}|t_{\mu\alpha}|^{2}|\mu\rangle\langle\mu|\int_{0}^{\pi}\frac{\sin^{2}(k_{\alpha})}{E-\epsilon_{k_{\alpha}\mu}+i0^{+}}d\arccos\left(\frac{\epsilon_{k_{\alpha}\mu}-w_{\alpha}}{2t_{\alpha}}\right)\nonumber \\
= & \sum_{\mu}\frac{|t_{\mu\alpha}|^{2}}{2t_{\alpha}}|\mu\rangle\langle\mu|\frac{1}{\pi}\int_{w_{\alpha}-2t_{\alpha}}^{w_{\alpha}+2t_{\alpha}}\frac{\sin^{2}\left[\arccos\left(\frac{\epsilon_{k_{\alpha}\mu}-w_{\alpha}}{2t_{\alpha}}\right)\right]}{\sqrt{1-\left(\frac{\epsilon_{k_{\alpha}\mu}-w_{\alpha}}{2t_{\alpha}}\right)^{2}}}\frac{1}{E-\epsilon_{k_{\alpha}\mu}+i0^{+}}d\epsilon_{k_{\alpha}\mu}\nonumber \\
= & \frac{|t_{\mu\alpha}|^{2}}{2t_{\alpha}}\left[x+\sqrt{\frac{x+i0^{+}+1}{x+i0^{+}-1}}-(x+i0^{+})\sqrt{\frac{x+i0^{+}+1}{x+i0^{+}-1}}\right],
\end{alignat}
where $x=\frac{E-w_{\alpha}}{2t_{\alpha}}$. In Fig. 1, we plot the
real part (red solid line) and imaginary part (blue dashed line) of
$2t_{\alpha}\Sigma_{\alpha}/|t_{\mu\alpha}|^{2}$ versus $\frac{E-w_{\alpha}}{2t_{\alpha}}$.
We observe that in the wide-band limit $\frac{E-w_{\alpha}}{2t_{\alpha}}\rightarrow0$,
the imaginary part of $2t_{\alpha}\Sigma_{\alpha}/|t_{\mu\alpha}|^{2}$
approaches one and its real part vanishes. Therefore, we can safely
ignore the level shift of the central region caused by the real part
of self-energy, retaining only the imaginary part of self-energy.
Therefore in the wide-band limit, the self-energy of lead $\alpha$
($=L,R$) becomes: 
\begin{equation}
\Sigma_{\alpha}^{w}=-\frac{i}{2}\sum_{\mu}\frac{|t_{\mu\alpha}|^{2}}{t_{\alpha}}|\mu\rangle\langle\mu|=-\frac{i}{2}\Gamma_{\alpha}^{w}.\label{eq:self-energy-wide-band}
\end{equation}
It is also energy independent and proportional to the level width
function. 

To summarize, for semi-infinite tight-binding leads coupled locally
to the central system, we find under wide-band limit the following
expressions for level width functions, Green's functions, spectral
functions and density of states: 
\begin{alignat}{1}
\Gamma_{L}^{w}= & \sum_{\mu=0}^{N-1}\frac{|t_{\mu L}|^{2}}{t_{L}}|\mu\rangle\langle\mu|,\nonumber \\
\Gamma_{R}^{w}= & \sum_{\mu=0}^{N-1}\frac{|t_{\mu R}|^{2}}{t_{R}}|\mu\rangle\langle\mu|,\nonumber \\
G_{CC}^{w}(E)= & \frac{1}{E-H_{CC}+\frac{i}{2}\left(\Gamma_{L}^{w}+\Gamma_{R}^{w}\right)},\nonumber \\
G_{CC}^{w\dagger}(E)= & \frac{1}{E-H_{CC}-\frac{i}{2}\left(\Gamma_{L}^{w}+\Gamma_{R}^{w}\right)},\nonumber \\
A_{C}^{w}(E)= & 2\pi\varrho_{C}^{w}(E)=i{\rm Tr}[G_{CC}^{w}(E)-G_{CC}^{w\dagger}(E)].\label{eq:collections}
\end{alignat}
These functions are enough for us to determine the transmission coefficient
$T(E)$ and current $I$ in wide-band limit, provided that the Hamiltonian
of the central region $H_{CC}$ is specified. Before doing that, we
will introduce a further simplification by taking the zero temperature,
zero bias limit of Eq. (\ref{eq:Landauer}). In this limit, we obtain
in the following section the linear conductance of the central system
as a response to perturbed electron distributions of the lead caused
by a small bias voltage.

\section{Linear conductance at zero temperature}

Let's consider the case in which a small bias voltage $V$ is applied
to the left lead, making its chemical potential $\mu_{L}$ slightly
differ from the chemical potential of the right lead:
\begin{equation}
-eV=\mu_{L}-\mu_{R},
\end{equation}
where $-e$ is the charge of electron. When $V$ is small enough,
the difference of Fermi functions of the two leads is given by
\begin{equation}
f_{L}(E)-f_{R}(E)=f_{R}(E+eV)-f_{R}(E)\approx eV\frac{\partial f_{R}(E)}{\partial E}.
\end{equation}
At zero temperature, only states below Fermi energy $E_{F}$ are filled.
So the Fermi function $f_{R}(E)=\theta(E_{F}-E)$ and its derivative
$\frac{\partial f_{R}(E)}{\partial E}=-\delta(E_{F}-E)$. So at zero
temperature in low bias regime, the current due to Landauer formula
(\ref{eq:Landauer}) is given by: 
\begin{equation}
I_{0}=-\frac{eV}{h}T(E_{F}),\label{eq:linear-current}
\end{equation}
which is proportional to the bias voltage (thus a linear response)
and the transmission coefficient $T$ evaluated at the Fermi energy
$E_{F}$. The corresponding \textbf{electrical conductance} is then
given by
\begin{equation}
{\cal C}_{0}=\frac{d(-eI_{0})}{dV}=\frac{e^{2}}{h}T(E_{F}),\label{eq:linear-conductance}
\end{equation}
which is just the transmission coefficient multiplied by the conductance
quanta $\frac{e^{2}}{h}$. Prefect transmission $T(E_{F})=1$ at the
Fermi energy results in a quantized conductance $\frac{e^{2}}{h}$.
Note here that the spin degrees of freedom of incoming electrons have
been ignored. For spinful incoming electrons, the results for the
current and conductance will be twice of ${\cal C}_{0}$. Eq. (\ref{eq:linear-conductance})
may be understood as a microscopic version of \textbf{Ohm's law}.
The physical meaning of conductance ${\cal C}_{0}$ can be interpreted
as a linear response of the central system to a perturbation of the
lead's chemical potential. Therefore it is a \textbf{linear conductance}.
If the perturbation $-eV$ makes $\mu_{L}$ slightly higher then $\mu_{R}$,
a quantized ${\cal C}_{0}$ will count the number of available transport
channels from left to right lead. On the contrary, if the perturbation
makes $\mu_{L}$ slightly lower then $\mu_{R}$, a quantized ${\cal C}_{0}$
will be equal to the number of available transport channels from right
to left lead. If all transport channels at $E_{F}$ are chiral and
spatially separated, a quantized response ${\cal C}_{0}$ under a
small bias voltage will only count the number of transport channels
with correct chirality \emph{locally}. Therefore, in the Chern insulator's
edge state transport calculation, the conductance ${\cal C}_{0}$\textbf{
}is equal to \textbf{gap chirality} instead of number of edge modes
on the Fermi surface. 

The conductance ${\cal C}_{0}$ can also be derived directly from
\textbf{Kubo formula} of linear response theory. Interested readers
can consult Chapter 2 of Ref.\cite{VentraBook} or Chapter 7 of Ref.\cite{BruusBook}.

In the following section, we will calculate the spectral function
and linear conductance of two explicit models in the wide-band limit.
These model calculations highlight the effects of \textbf{resonance}
and \textbf{coherence} on electrical transport in noninteracting quantum
systems.

\section{Steady state current in a quantum dot: resonance and coherence}

In this section, we will apply the Landauer formula for linear conductance
to two explicitly examples: a single-level quantum dot and a double-level
quantum dot. In both examples, the spectral function and transmission
coefficient of the quantum dot can be obtained analytically in the
wide-band limit. These models serve as the starting point for study
of coherent transport in discrete level quantum systems. Some key
concepts in the resonant transport can be easily demonstrated in the
following calculations.

\subsection{Single-level quantum dot: spectral function and transmission coefficient}

The Hamiltonian of a single-level quantum dot is given by
\begin{equation}
H_{CC}=\epsilon_{0}|0\rangle\langle0|,
\end{equation}
where $\epsilon_{0}$ is the energy of level $|0\rangle$. A sketch
of the model is shown in Fig. 2. In the wide-band limit, we find from
Eq. (\ref{eq:collections}) the following expressions for level-width
functions and Green's functions: 
\begin{alignat}{1}
\Gamma_{L}^{w}= & \gamma_{0L}|0\rangle\langle0|,\\
\Gamma_{R}^{w}= & \gamma_{0R}|0\rangle\langle0|,\\
G_{CC}^{w}(E)= & \frac{1}{E-\epsilon_{0}+\frac{i}{2}\gamma_{0}}|0\rangle\langle0|,\\
G_{CC}^{w\dagger}(E)= & \frac{1}{E-\epsilon_{0}-\frac{i}{2}\gamma_{0}}|0\rangle\langle0|,
\end{alignat}
where we have introduced simplified notations $\gamma_{0\alpha}\equiv\frac{|t_{0\alpha}|^{2}}{t_{\alpha}}$
for $\alpha=L,R$ and $\gamma_{0}\equiv\gamma_{0L}+\gamma_{0R}$.
The spectral function and transmission coefficient are then given
by
\begin{alignat}{1}
A_{C}^{w}(E)= & i{\rm Tr}[G_{CC}^{w}(E)-G_{CC}^{w\dagger}(E)]=\frac{\gamma_{0}}{(E-\epsilon_{0})^{2}+\frac{1}{4}\gamma_{0}^{2}}=2\pi\varrho_{C}^{w}(E),\\
T^{w}(E)= & {\rm Tr}\left[G_{CC}^{w\dagger}(E)\Gamma_{R}^{w}(E)G_{CC}^{w}(E)\Gamma_{L}^{w}(E)\right]=\frac{\gamma_{0L}\gamma_{0R}}{(E-\epsilon_{0})^{2}+\frac{1}{4}\gamma_{0}^{2}}.
\end{alignat}
We see that both of them are of \textbf{Lorentzian shape}. In zero
temperature, low bias regime, the density of states and linear conductance
are determined by states at the Fermi energy:
\begin{alignat}{1}
\varrho_{C}^{w}(E_{F})= & \frac{1}{2\pi}\frac{\gamma_{0}}{(E_{F}-\epsilon_{0})^{2}+\frac{1}{4}\gamma_{0}^{2}},\\
{\cal C}_{0}= & \frac{e^{2}}{h}\frac{\gamma_{0L}\gamma_{0R}}{(E_{F}-\epsilon_{0})^{2}+\frac{1}{4}\gamma_{0}^{2}}=\frac{e^{2}}{\hbar}\frac{\gamma_{0L}\gamma_{0R}}{\gamma_{0L}+\gamma_{0R}}\varrho_{C}^{w}(E_{F}).
\end{alignat}
The maximum of linear conductance is reached when the energy of an
incoming electron matches the energy of the quantum dot $\epsilon_{0}$.
In this case, we obtain the \textbf{resonant transmission coefficient}:
\begin{equation}
T^{w}(E)=\frac{4\gamma_{0L}\gamma_{0R}}{(\gamma_{0L}+\gamma_{0R})^{2}},
\end{equation}
which is always equal to $1$ if the dot-lead coupling is symmetric
($\gamma_{0L}=\gamma_{0R}$), no matter how strong (or weak) of the
coupling is.

\subsection{Double-level quantum dot: spectral function and transmission coefficient}

Let's now apply the Landauer formula (\ref{eq:Landauer}) to the electrical
transport through a two-level quantum dot. The Hamiltonian of the
quantum dot is given by
\begin{equation}
H_{CC}=\epsilon_{0}|0\rangle\langle0|+\epsilon_{1}|1\rangle\langle1|+t_{c}(|0\rangle\langle1|+|1\rangle\langle0|)=\begin{bmatrix}\epsilon_{0} & t_{c}\\
t_{c} & \epsilon_{1}
\end{bmatrix}.
\end{equation}
where $\epsilon_{0,1}$ are energies of decoupled levels $|0\rangle,|1\rangle$
and $t_{c}$ is the coupling strength between the two levels. In the
wide-band limit, we find from Eq. (\ref{eq:collections}) the following
expressions for level-width function and Green's function:
\begin{equation}
\Gamma_{L}^{w}=\begin{bmatrix}\frac{|t_{0L}|^{2}}{t_{L}} & 0\\
0 & \frac{|t_{1L}|^{2}}{t_{L}}
\end{bmatrix}\qquad\Gamma_{R}^{w}=\begin{bmatrix}\frac{|t_{0R}|^{2}}{t_{R}} & 0\\
0 & \frac{|t_{1R}|^{2}}{t_{R}}
\end{bmatrix},
\end{equation}
\begin{alignat}{1}
G_{CC}^{w}(E)= & \frac{1}{(E-\epsilon_{0}+i\gamma_{0})(E-\epsilon_{1}+i\gamma_{1})-t_{c}^{2}}\begin{bmatrix}E-\epsilon_{1}+i\gamma_{1} & t_{c}\\
t_{c} & E-\epsilon_{0}+i\gamma_{0}
\end{bmatrix},\\
G_{CC}^{w\dagger}(E)= & \frac{1}{(E-\epsilon_{0}-i\gamma_{0})(E-\epsilon_{1}-i\gamma_{1})-t_{c}^{2}}\begin{bmatrix}E-\epsilon_{1}-i\gamma_{1} & t_{c}\\
t_{c} & E-\epsilon_{0}-i\gamma_{0}
\end{bmatrix},
\end{alignat}
where we have introduced the simplified notations $\gamma_{\mu}=\frac{|t_{\mu L}|^{2}}{2t_{L}}+\frac{|t_{\mu R}|^{2}}{2t_{R}}$
for $\mu=0,1$. The spectral function and transmission coefficient
are then given by
\begin{equation}
A_{C}^{w}(E)=-2{\rm Im}{\rm Tr}[G_{CC}^{w}(E)]=-2{\rm Im}\left[\frac{(E-\epsilon_{0}+i\gamma_{0})+(E-\epsilon_{1}+i\gamma_{1})}{(E-\epsilon_{0}+i\gamma_{0})(E-\epsilon_{1}+i\gamma_{1})-t_{c}^{2}}\right],
\end{equation}
\begin{alignat}{1}
T^{w}(E)= & \frac{1}{|(E-\epsilon_{0}+i\gamma_{0})(E-\epsilon_{1}+i\gamma_{1})-t_{c}^{2}|^{2}}\nonumber \\
\times & {\rm Tr}\left(\begin{bmatrix}E-\epsilon_{1}-i\gamma_{1} & t_{c}\\
t_{c} & E-\epsilon_{0}-i\gamma_{0}
\end{bmatrix}\begin{bmatrix}\frac{|t_{0R}|^{2}}{t_{R}} & 0\\
0 & \frac{|t_{1R}|^{2}}{t_{R}}
\end{bmatrix}\begin{bmatrix}E-\epsilon_{1}+i\gamma_{1} & t_{c}\\
t_{c} & E-\epsilon_{0}+i\gamma_{0}
\end{bmatrix}\begin{bmatrix}\frac{|t_{0L}|^{2}}{t_{L}} & 0\\
0 & \frac{|t_{1L}|^{2}}{t_{L}}
\end{bmatrix}\right).
\end{alignat}

To simplify these expressions, we focus on a minimum extension of
single-level quantum dot, i.e., only one level of the two-level quantum
dot is coupled to the lead. Our LCR system then forms a ``\textbf{T-junction}''
as shown in Fig. 3. In this case, we may set $t_{1L}=t_{1R}=0$, which
also results in $\gamma_{1}=0$. Furthermore, we consider a symmetric
dot-lead coupling by setting $\frac{|t_{0L}|^{2}}{t_{L}}=\frac{|t_{0R}|^{2}}{t_{R}}=\gamma_{0}$.
With these simplifications, we obtain the following expressions for
the spectral function $A_{C}^{w}(E)$ and transmission coefficient
$T^{w}(E)$ of the quantum dot: 
\begin{alignat}{1}
A_{C}^{w}(E)= & 2\gamma_{0}\frac{1+\frac{t_{c}^{2}}{(E-\epsilon_{1})^{2}}}{\left(E-\epsilon_{0}-\frac{t_{c}^{2}}{E-\epsilon_{1}}\right)^{2}+\gamma_{0}^{2}},\\
T^{w}(E)= & \frac{\gamma_{0}^{2}}{\left(E-\epsilon_{0}-\frac{t_{c}^{2}}{E-\epsilon_{1}}\right)^{2}+\gamma_{0}^{2}}.
\end{alignat}
We see that none of these functions are of Lorentzian shape. Moreover,
the transmission $T^{w}(E)=0$ if $E=\epsilon_{1}$. This is an \textbf{anti-resonance}
effect. Roughly speaking, an incoming electron has two possible paths
to go through the ``T-junction'' quantum dot. We denote them by
our localized basis as (i) $|-1,0\rangle\rightarrow|0\rangle\rightarrow|1,0\rangle$
and (ii) $|-1,0\rangle\rightarrow|0\rangle\rightarrow|1\rangle\rightarrow|0\rangle\rightarrow|1,0\rangle$.
Along path (i), the incident electron will reach level $\epsilon_{0}$
and then directly go through the quantum dot. Along path (ii), it
will first hop from level $\epsilon_{0}$ to $\epsilon_{1}$ through
the coupling $t_{c}$, and then hop back to the level $\epsilon_{0}$.
If the energy $E$ of the incoming electron matches $\epsilon_{1}$,
waves separately following the two paths in the quantum dot will have
a $\pi$ phase difference when they meet with each other at the level
$\epsilon_{0}$, resulting in a \textbf{destructive interference}.
Therefore, under the condition $E=\epsilon_{1}$, we have a \textbf{coherence-induced
destruction of tunneling} in this simple example. In Fig. 4, the spectral
and transmission functions for three different values of the coupling
strength $t_{c}$ between two levels of the quantum dot are shown.
In all cases, we observe clear collapses of the transmission coefficient
at $E=\epsilon_{1}$. 

In low bias regime at zero temperature, the density of states and
linear conductance of this model are also determined by states at
the Fermi energy:
\begin{alignat}{1}
\varrho_{C}^{w}(E_{F})= & \frac{\gamma_{0}}{\pi}\frac{1+\frac{t_{c}^{2}}{(E_{F}-\epsilon_{1})^{2}}}{\left(E_{F}-\epsilon_{0}-\frac{t_{c}^{2}}{E_{F}-\epsilon_{1}}\right)^{2}+\gamma_{0}^{2}},\\
{\cal C}_{0}= & \frac{e^{2}}{h}\frac{\gamma_{0}^{2}}{\left(E_{F}-\epsilon_{0}-\frac{t_{c}^{2}}{E_{F}-\epsilon_{1}}\right)^{2}+\gamma_{0}^{2}}=\frac{e^{2}}{\hbar}\frac{1}{2}\frac{\gamma_{0}}{1+\frac{t_{c}^{2}}{(E_{F}-\epsilon_{1})^{2}}}\varrho_{C}^{w}(E_{F}).
\end{alignat}
With respect to $E_{F}$, both of them have two peaks in quite different
shapes, originating from nontrivial effects of quantum coherence in
the system.

\section{Current across a tight-binding chain: metal-insulator transitions}

The formulas we summarized in the collection of Eq. (\ref{eq:collections})
are also appropriate for the study of linear transport in a tight-binding
chain. In this case, we make the following choices for level-width
function (in wide-band limit):
\begin{equation}
\Gamma_{L}^{w}=\frac{|t_{CL}|^{2}}{t_{L}}|0\rangle\langle0|,\qquad\Gamma_{R}^{w}=\frac{|t_{CR}|^{2}}{t_{R}}|N-1\rangle\langle N-1|,
\end{equation}
that is, the left (right) lead is only coupled to the first (last)
site of the central system. The Hamiltonian $H_{CC}$ describing the
system of our interest will be chosen to have the following form:
\begin{equation}
H_{CC}=\sum_{\mu=0}^{N-1}\epsilon_{\mu}|\mu\rangle\langle\mu|+t_{C}\sum_{\mu=0}^{N-2}\left(|\mu\rangle\langle\mu+1|+{\rm h.c.}\right),
\end{equation}
which represents a typical tight-binding chain with onsite potential
$\{\epsilon_{\mu}|\mu=0,...,N-1\}$ and nearest neighbor hopping $t_{C}$
(Other situations may include long-range, complex and site-dependent
hopping amplitudes. But it should be straightforward to generalize
the treatment here to those more complicated cases). The configuration
of this LRC system is illustrated in Fig. 5. The retarded Green's
function, spectral function and transmission coefficient of the tight-binding
chain are the computed by the following formulas:
\begin{alignat}{1}
G_{CC}^{w}(E)= & \frac{1}{E-H_{CC}+\frac{i}{2}\left(\gamma_{L}|0\rangle\langle0|+\gamma_{R}|N-1\rangle\langle N-1|\right)},\\
A_{C}^{w}(E)= & i{\rm Tr}[G_{CC}^{w}(E)-G_{CC}^{w\dagger}(E)],\\
T^{w}(E)= & \gamma_{L}\gamma_{R}|\langle N-1|G_{CC}^{w}(E)|0\rangle|^{2}.
\end{alignat}
where $\gamma_{\alpha}=\frac{|t_{C\alpha}|^{2}}{t_{\alpha}}$ for
$\alpha=L,R$. The last equality is obtained from Eq. (\ref{eq:Transcoeff}).
As a useful observation, the transmission property of $H_{CC}$ at
a given energy $E$ is simply determined by a single corner matrix
element $\langle N-1|G_{CC}^{w}(E)|0\rangle$ of the retarded Green's
function. 

To give an explicit demonstration, we consider the onsite potential
of $H_{CC}$ to have the following expression:
\begin{equation}
\epsilon_{\mu}=\epsilon\cos\left(2\pi\frac{p}{q}\mu-k_{y}\right)\qquad\mu=0,...,N-1\qquad p,q\in\mathbb{Z}.
\end{equation}
Here $k_{y}\in[0,2\pi)$ is a phase shift of the onsite potential,
which will be given a concrete physical meaning in our next example.
With this choice, the Hamiltonian of our central system is explicitly
given by: 
\begin{equation}
H_{{\rm AAH}}=\epsilon\sum_{\mu=0}^{N-1}\cos\left(2\pi\frac{p}{q}\mu-k_{y}\right)|\mu\rangle\langle\mu|+t_{C}\sum_{\mu=0}^{N-2}\left(|\mu\rangle\langle\mu+1|+{\rm h.c.}\right).
\end{equation}
It is usually called \textbf{Aubry-André-Harper (AAH) model}, which
has been thoroughly explored in the study of metal-insulator transitions.
Its two-dimensional parent model, often called the \textbf{Hofstadter
model}, is also a popular prototype in the study quantum Hall effects
and topological insulators. 

As discussed in previous sections, at zero temperature, the linear
transport properties of $H_{{\rm AAH}}$ is determined by its spectral
function $A_{C}^{w}(E)$ and transmission coefficient $T^{w}(E)$
evaluated at the Fermi energy $E_{F}$, giving us:
\begin{alignat}{1}
A_{C}^{w}(E_{F})= & i{\rm Tr}[G_{CC}^{w}(E_{F})-G_{CC}^{w\dagger}(E_{F})],\\
T^{w}(E_{F})= & \gamma_{L}\gamma_{R}|\langle N-1|G_{CC}^{w}(E_{F})|0\rangle|^{2}.
\end{alignat}
In Fig. 6, we show the numerical results of these two functions with
respect to $E_{F}$ and $k_{y}$ for two typical cases. There are
third points to mention. First, in the plots for spectral function
$A_{C}^{w}(E_{F})$, the states traversing the gaps are chiral edge
modes. These modes are localized at the edges of the AAH chain and
therefore cannot contribute to the transport, as reflected in the
lower panels for the transmission coefficient $T^{w}(E_{F})$. Second,
there is a metal-insulator phase transition at $\epsilon=2t_{C}$
for an irrational $\alpha$. When $\epsilon<2t_{C}$ ($\epsilon>2t_{C}$),
the bulk states (bright regions) separated by gaps in $A_{C}^{w}(E_{F})$
are conducting (insulating) with non-vanishing (vanishing) transmission
coefficients $T^{w}(E_{F})$. Third, resolutions of edge mode in spectral
function are better for weaker couplings. Since the couplings are
introduced at the boundary of the AAH chain, it is expected that a
strong system-lead coupling will make the edge states less observable.
On the contrary, to get better transmission properties, we need to
choose couplings $\gamma_{L,R}$ closer to the energy range of the
central system.

\section{Current along the edge of a two-dimensional lattice: topologically
quantized transport}

As a final example of these notes, we will apply the Landauer formula
to compute the conductance of a two-dimensional lattice. We choose
the \textbf{Hofstadter model} (the parent of AAH model) as the central
system of our interest. It describes noninteracting electrons hopping
on a two-dimensional square lattice in a perpendicular magnetic field.
In real space, the Hamiltonian is given by:
\begin{equation}
H_{CC}=t_{x}\sum_{x=0}^{N_{x}-2}\sum_{y=0}^{N_{y}-1}\left(|x\rangle\langle x+1|\otimes|y\rangle\langle y|+{\rm h.c.}\right)+t_{y}\sum_{x=0}^{N_{x}-1}\sum_{y=0}^{N_{y}-2}\left(e^{-i2\pi\frac{p}{q}x}|x\rangle\langle x|\otimes|y\rangle\langle y+1|+{\rm h.c.}\right),\label{eq:Hofstadter}
\end{equation}
where $t_{x}$ ($t_{y}$) is the nearest neighbor hopping amplitude
along $x$ ($y$) direction of the lattice. The system contains $(N_{x}+1)\times(N_{y}+1)$
lattice sites. This model can be reduced to the AAH model if we take
periodic boundary conditions along $y$-direction, and interpreting
the parameter $k_{y}$ in the AAH model as the quasimomentum along
$y$-direction. 

To evaluate the current, we now couple the central system described
by Hamiltonain (\ref{eq:Hofstadter}) to leads. On the left hand side
of $H_{CC}$, each lattice site $(0,y)$ for $y=0,1,...,N_{y}-1$
is coupled to a semi-infinite tight-binding chain extends from $(-\infty,y)$
to $(-1,y)$. On the right hand side of $H_{CC}$, each lattice site
$(N_{x}-1,y)$ for $y=0,1,...,N_{y}-1$ is also coupled to a semi-infinite
tight-binding chain extends from $(N_{x},y)$ to $(+\infty,y)$. The
configuration of this LRC system is illustrated in Fig. 7. In the
wide-band limit, the left and right level-width functions are given
by:
\begin{alignat}{1}
\Gamma_{L}^{w}= & |0\rangle\langle0|\otimes\sum_{y=0}^{N_{y}-1}\gamma_{yL}|y\rangle\langle y|,\\
\Gamma_{R}^{w}= & |N_{x}-1\rangle\langle N_{x}-1|\otimes\sum_{y=0}^{N_{y}-1}\gamma_{yR}|y\rangle\langle y|,
\end{alignat}
where we took $\gamma_{y\alpha}=\frac{|t_{y\alpha}|^{2}}{t_{\alpha}}$
for $\alpha=L,R$. The Hamiltonian $H_{CC}$, together with level-width
functions $\Gamma_{L,R}^{w}$, determines the retarded Green's function,
spectral function and transmission coefficient of the central region:
\begin{equation}
G_{CC}^{w}(E)=\frac{1}{E-H_{CC}+\frac{i}{2}\sum_{y=0}^{N_{y}-1}\left(\gamma_{yL}|0\rangle\langle0|+\gamma_{yR}|N_{x}-1\rangle\langle N_{x}-1|\right)\otimes|y\rangle\langle y|},
\end{equation}
\begin{alignat}{1}
A_{C}^{w}(E)= & -2{\rm Im}\{{\rm Tr}[G_{CC}^{w}(E)]\},\\
T^{w}(E)= & {\rm Tr}\left[G_{CC}^{w\dagger}(E)\Gamma_{R}^{w}(E)G_{CC}^{w}(E)\Gamma_{L}^{w}(E)\right]\nonumber \\
= & \sum_{y,y'=0}^{N_{y}-1}\gamma_{yL}\gamma_{y'R}|\langle N_{x}-1,y'|G_{CC}^{w}(E)|0,y\rangle|^{2}.
\end{alignat}
The transmission coefficient $T^{w}(E)$ has a transparent interpretation
as summation over scattering amplitudes from left ($x=0$) to right
($x=N_{x}-1$) edges for all possible incoming and outgoing sites. 

As a demonstration of these formulas, we consider the transmission
coefficient of Hofstadter model at zero temperature in zero bias limit.
For $p/q=1/3$ and $p/q=1/5$, numerical results for $T^{w}(E_{F})$
vs. $E_{F}$ are shown in Fig. 8. In both cases, we observed quantized
transmission coefficients (and therefore conductances) in the spectral
gap of the system. As illustrated in the plots of spectral function.
current-carrying states in the gap are chiral edge modes localized
at the boundary of the two-dimensional lattice. The quantization of
edge state conductance in a spectral gap of the Hofstadter model is
topological, equaling to the summation of bulk band Chern numbers
below the considered gap. This bulk-edge correspondence principle
holds in general for noninteracting fermionic Chern insulators. Interested
readers can consult Ref.\cite{HatsugaiBEC} for more details.

\section{Summary and plan for future extensions}

In these notes, we give a brief introduction to matrix Green's function,
focusing on its applications to steady state transport in discrete
level quantum systems without many-body interactions. The Landauer
formula is derived from a naive scattering viewpoint, and then applied
to compute linear conductance of a single-level quantum dot, a double-level
quantum dot, a tight-binding chain and a two-dimensional tight-binding
lattice at zero temperature under wide-band limit. Even though a lot
of approximations and simplifications have been made, the results
still demonstrate the important roles of \emph{resonance, coherence,
disorder and topology} in quantum transport.

Further extensions of these notes may include the following topics:
\begin{itemize}
\item Steady state transport in periodically driven quantum systems: a combination
of Green's function and Floquet formalism
\item Initial states with inter-channel coherence
\item Effects of finite temperature and many-body interaction
\item Quantum pumping
\end{itemize}

\subsection*{Acknowledgments}

L. Zhou thanks Prof. Jian-Sheng Wang for his pedagogical lectures
on NEGF, and helpful discussions with Prof. Jiangbin Gong and Mr.
Han Hoe Yap, which motivated the preparation of these notes.

\newpage{}

\begin{figure}
\begin{centering}
\includegraphics[scale=0.6]{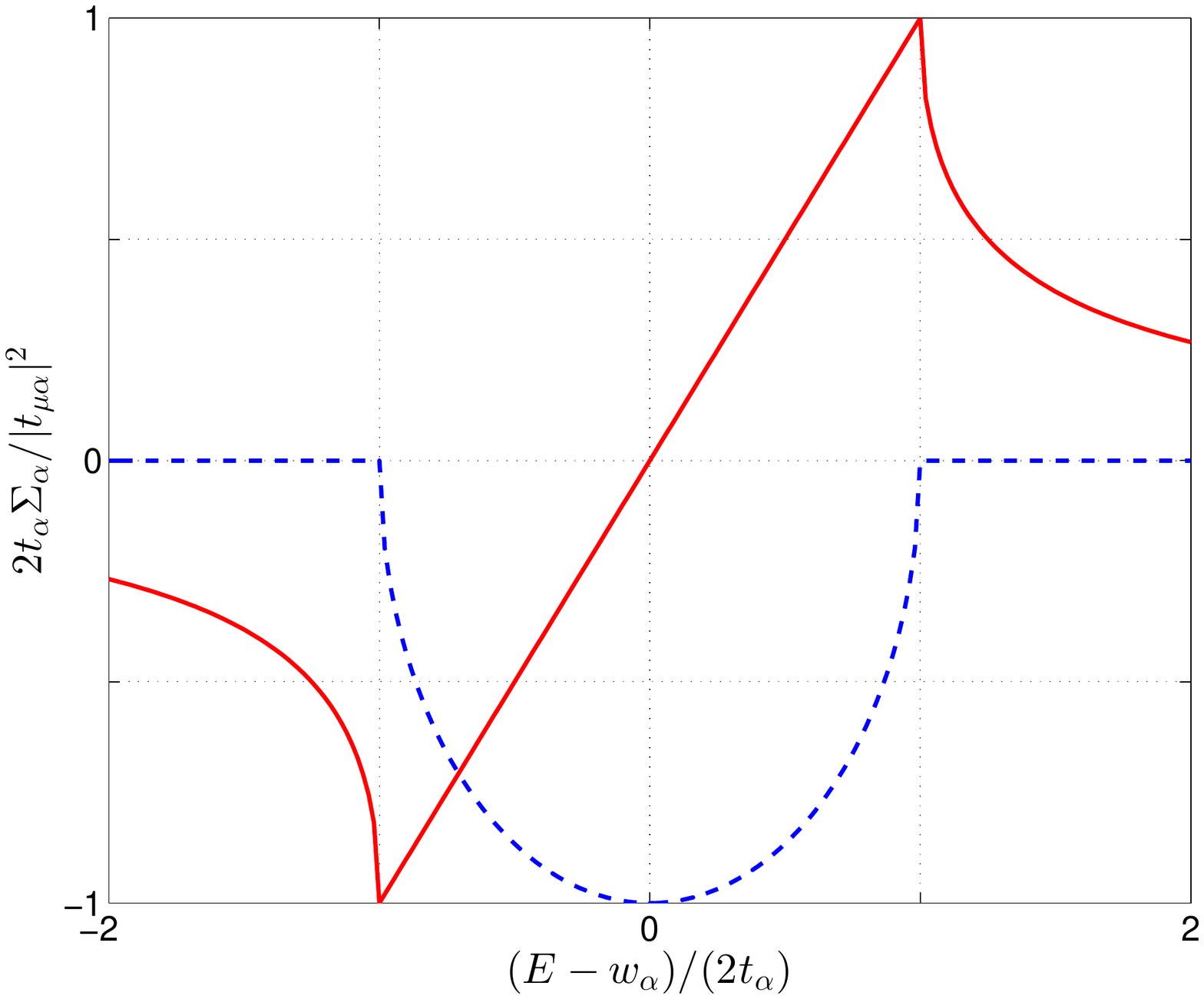}
\par\end{centering}

\caption{Real part (red solid line) and imaginary part (blue dashed line) of
the lead $\alpha$'s self-energy $\Sigma_{\alpha}$, plotted with
respect to the difference between the energy of incoming electron
$E$ and the band center $w_{\alpha}$ of lead $\alpha$. }
\end{figure}

\begin{figure}
\begin{centering}
\includegraphics[scale=0.6]{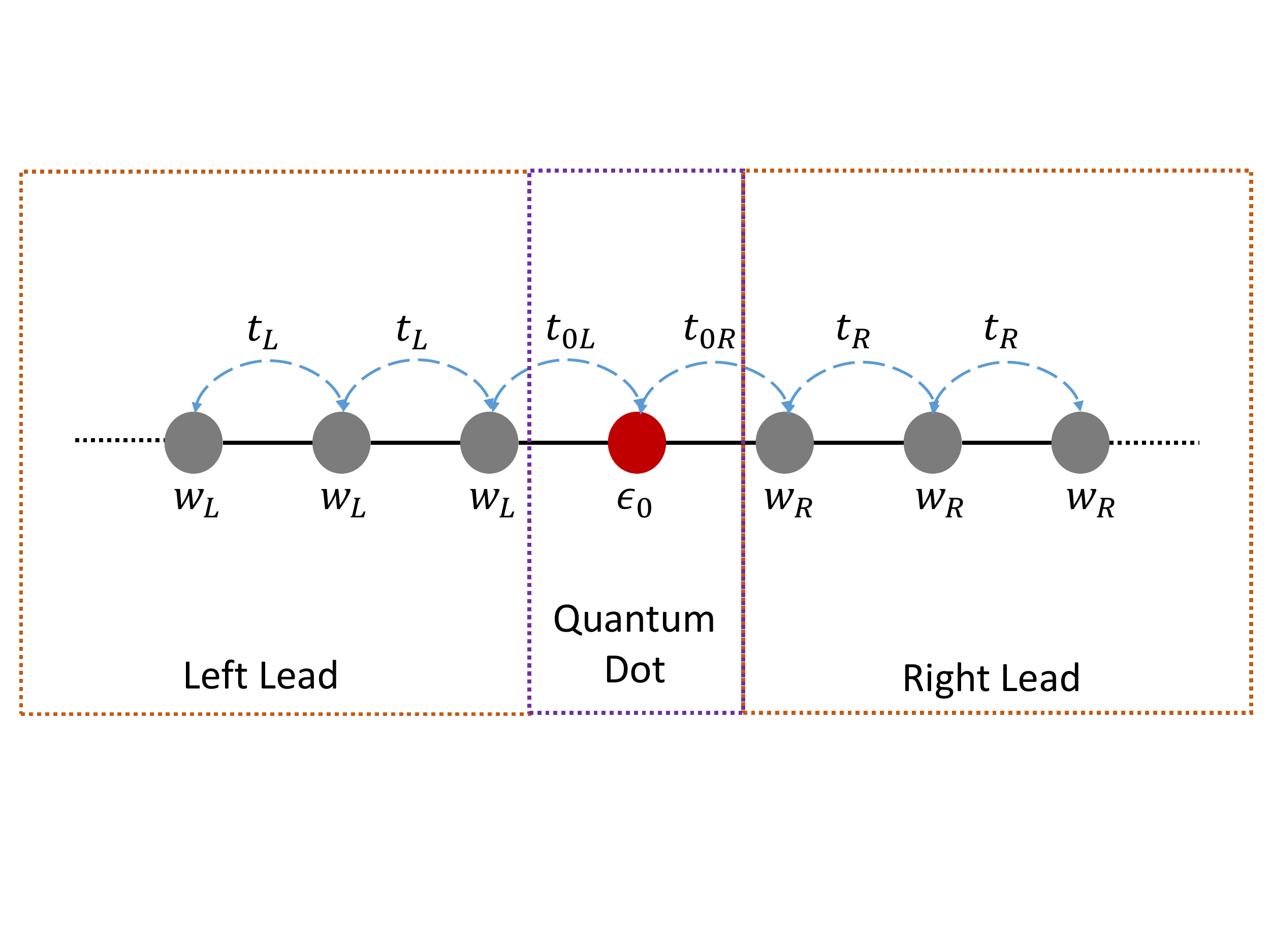}
\par\end{centering}

\caption{A single-level quantum dot coupled to two semi-infinite tight-binding
leads.}
\end{figure}

\begin{figure}
\begin{centering}
\includegraphics[scale=0.6]{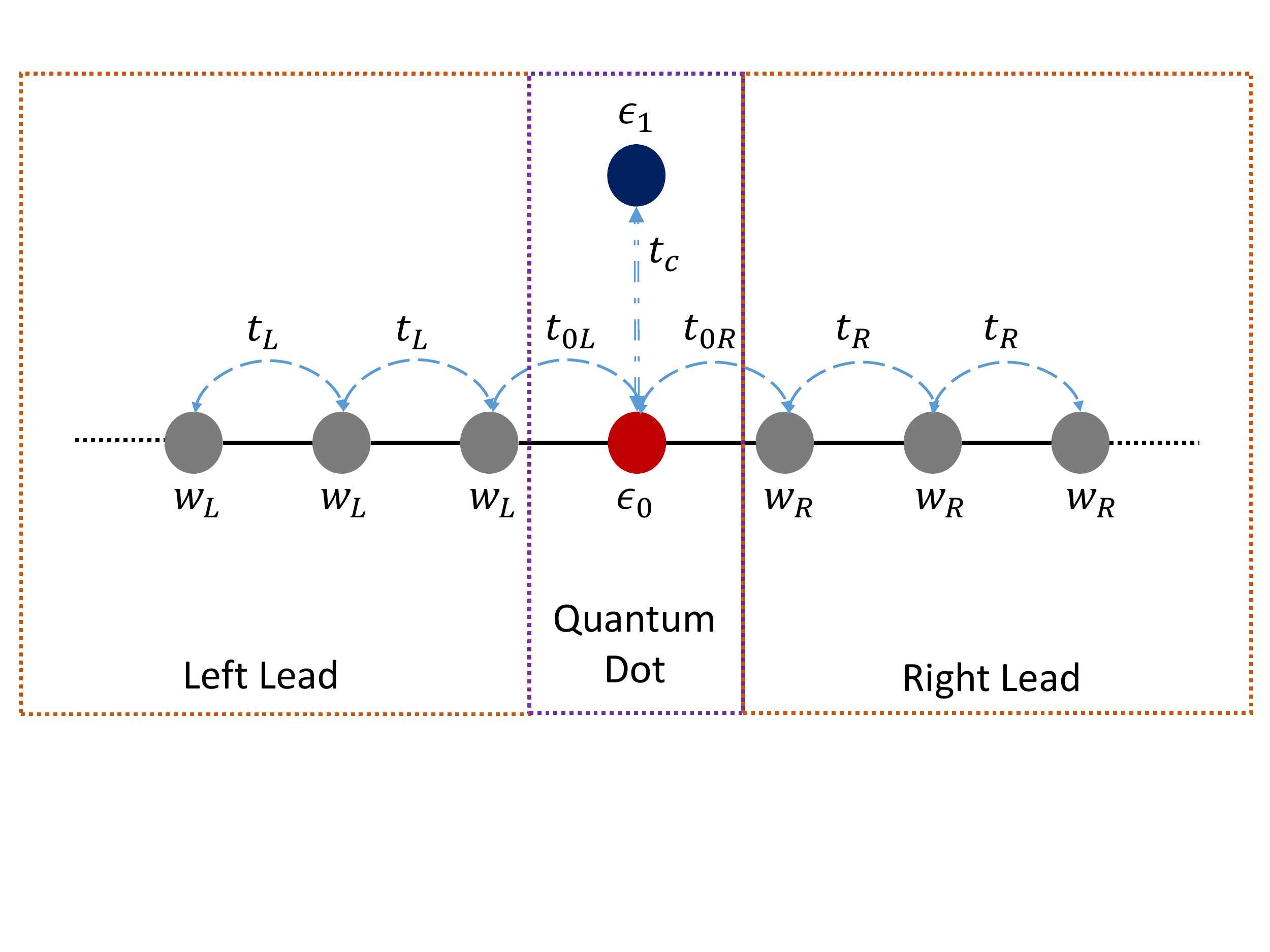}
\par\end{centering}

\caption{A double-level quantum dot coupled to semi-infinite tight-binding
leads. The leads only couple to one of the two levels, forming a T-shape
junction.}
\end{figure}

\begin{figure}
\begin{centering}
\includegraphics[scale=0.6]{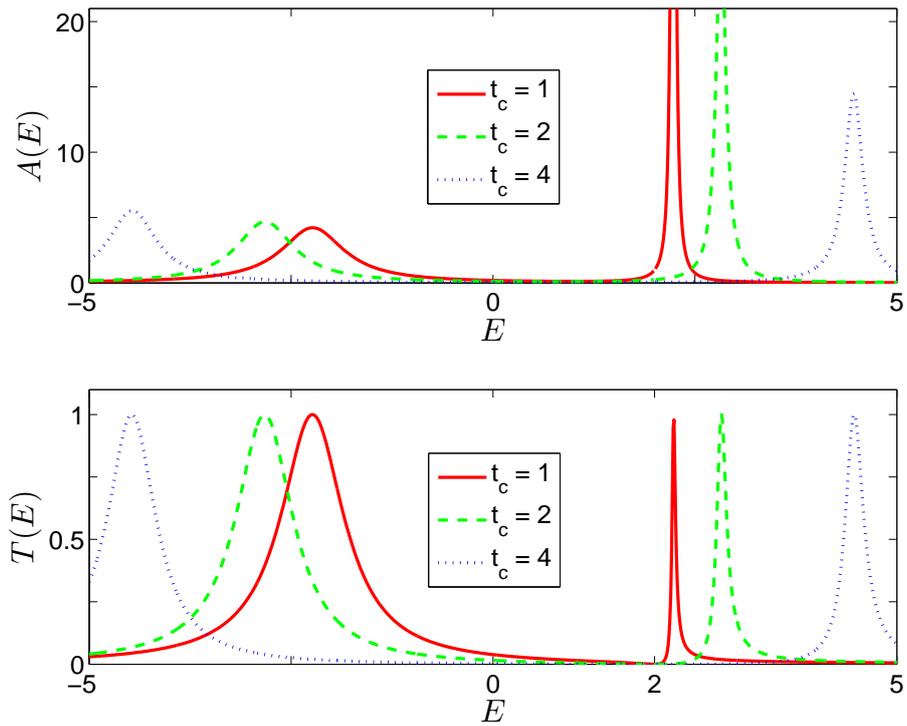}
\par\end{centering}

\caption{Spectral function $A(E)$ and transmission coefficient $T(E)$ of
the double-level quantum dot coupled to semi-infinite tight-binding
leads in wide-band limit. System parameters are $\epsilon_{0}=-2$,
$\epsilon_{1}=2$ and $\gamma_{0}=0.5$. Only level $\epsilon_{0}$
is coupled to the leads as shown in Fig. 3. The results for three
different coupling strengths between the two levels of the quantum
dot are shown.}
\end{figure}

\begin{figure}
\begin{centering}
\includegraphics[scale=0.6]{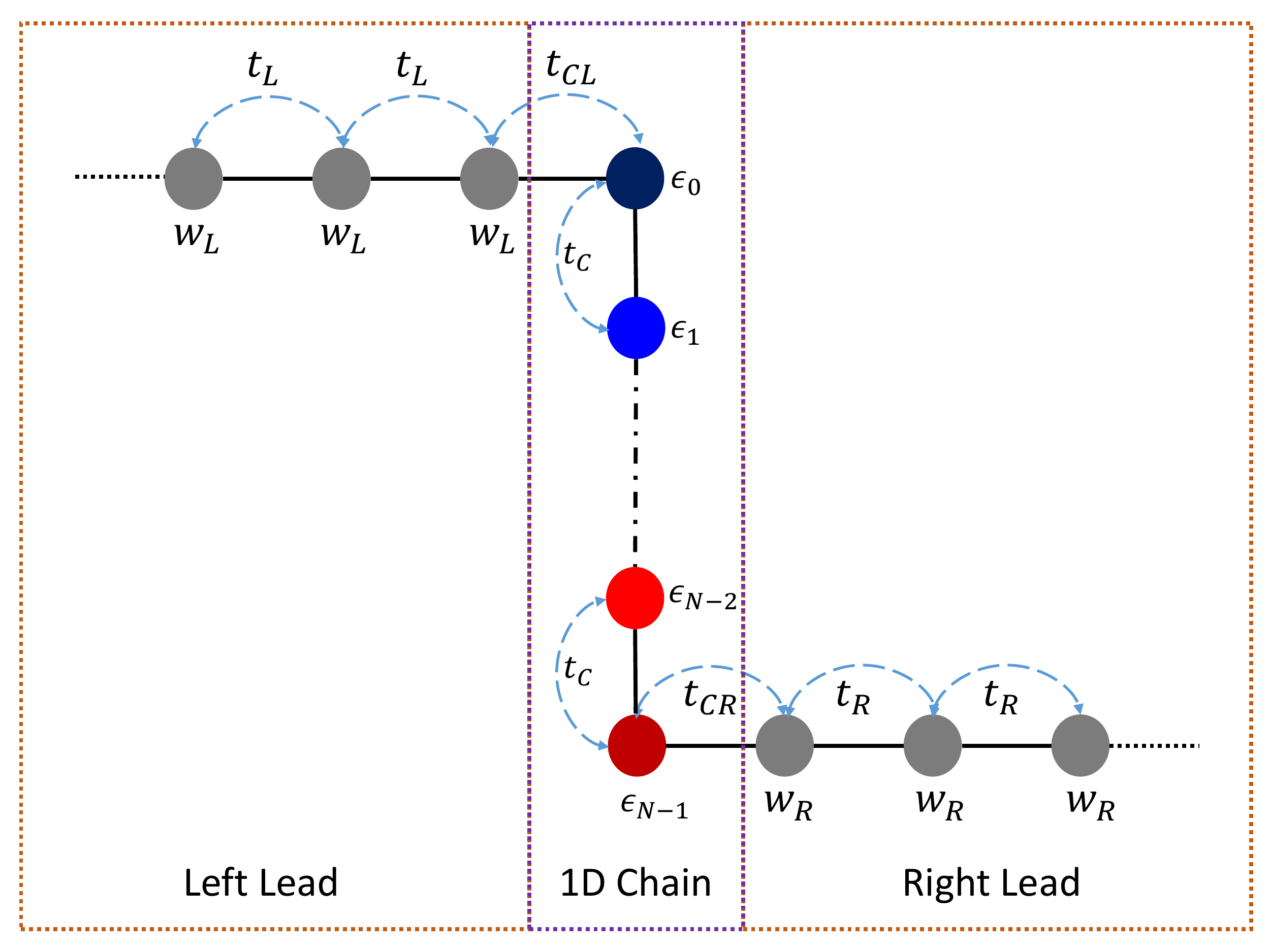}
\par\end{centering}

\caption{A tight-binding chain coupled to semi-infinite tight-binding leads
at its two boundary sites.}
\end{figure}

\begin{figure}
\begin{centering}
\includegraphics[scale=0.6]{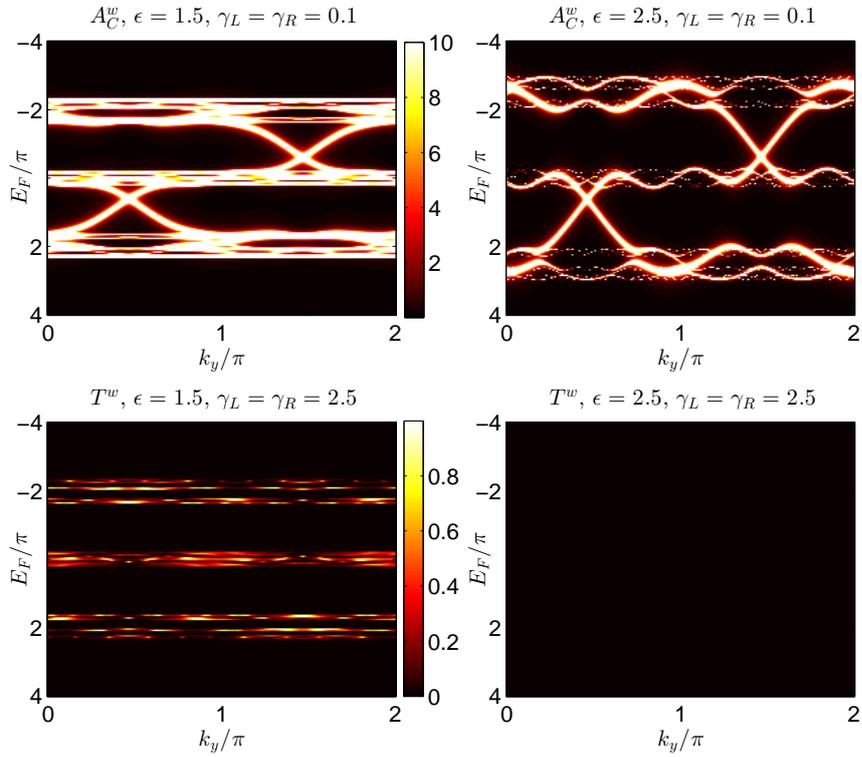}
\par\end{centering}

\caption{Spectral function $A_{C}^{w}(E_{F})$ and transmission coefficient
$T^{w}(E_{F})$ of the AAH model at zero temperature in zero bias,
wide-band limit. Results are shown for two different values of onsite
potential $\epsilon=1.5$ and $2.5$. For spectral function, we take
$\gamma_{L}=\gamma_{R}=0.1$. For transmission coefficient, we take
$\gamma_{L}=\gamma_{R}=2.5$. The other system parameters are $t_{C}=1$,
$\frac{p}{q}=\frac{\sqrt{5}-1}{2}$, and $N=59$. }
\end{figure}

\begin{figure}
\begin{centering}
\includegraphics[scale=0.6]{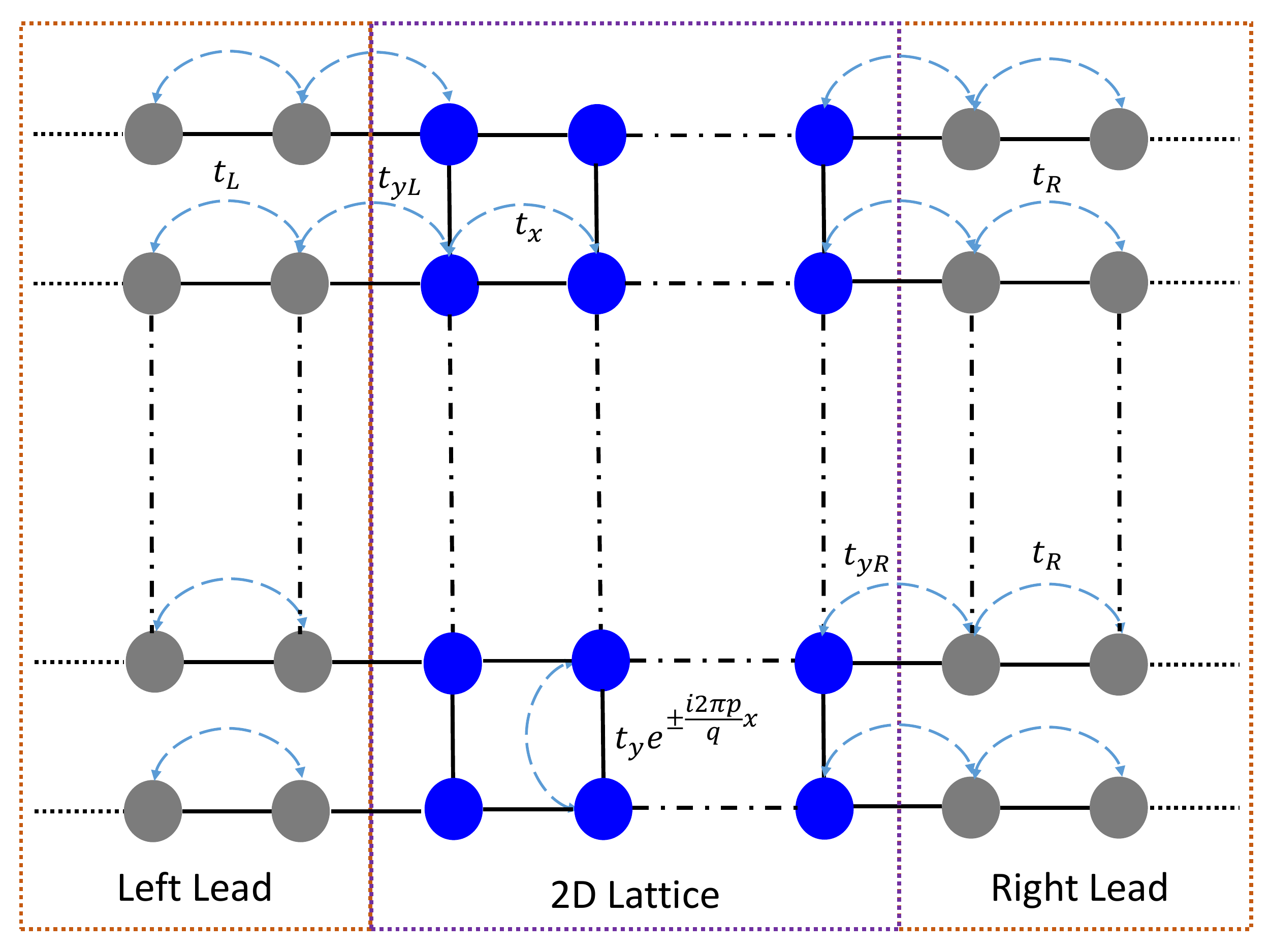}
\par\end{centering}

\caption{A two-dimensional lattice (Hofstadter model) coupled to semi-infinite
tight-binding leads at its left and right boundary sites.}
\end{figure}

\begin{figure}
\begin{centering}
\includegraphics[scale=0.6]{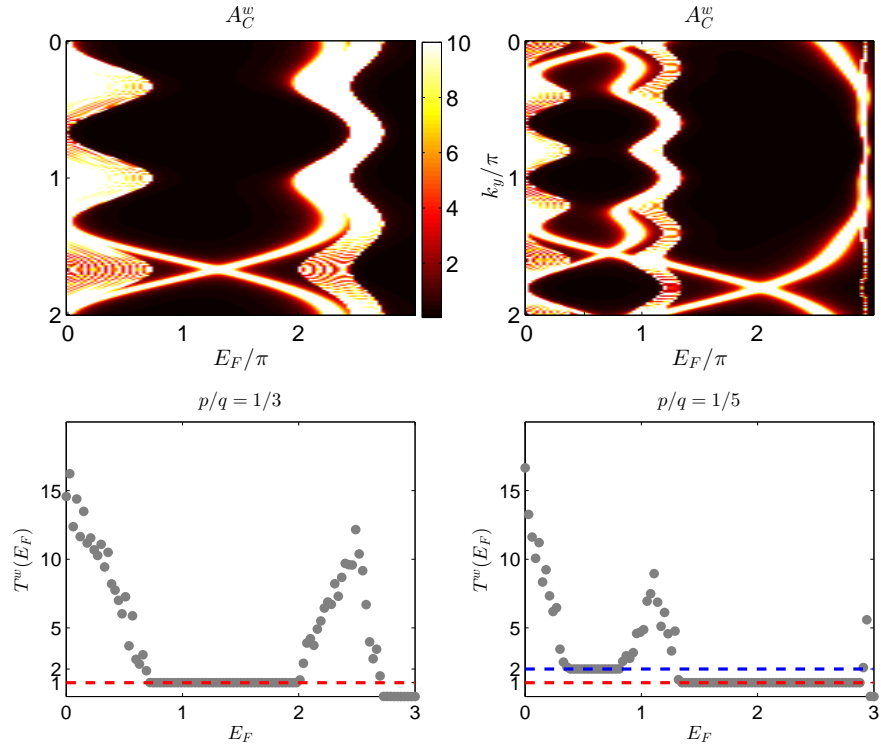}
\par\end{centering}

\caption{Transmission coefficients of the Hofstadter model at different $E_{F}$
for $p/q=1/3$ and $p/q=1/5$. The other system parameters are chosen
as $N_{x}=N_{y}=44$, $t_{x}=t_{y}=1$ and $\gamma_{yL}=\gamma_{yR}=1$.
The upper panels show spectral functions calculated for the two cases
with periodic boundary conditions along $y$-direction. Bright lines
traversing the spectral gaps represent chiral edge modes. The lower
panels show numerically evaluated transmission coefficients vs. Fermi
energy $E_{F}$ (in gray dots). Dashed lines are guide to the eye
to point out the quantization of transmission coefficient (and conductance)
in the spectral gap. }
\end{figure}

\end{document}